# Room temperature optical control of spin states in organic diradicals

Rituparno Chowdhury[1,†], Alistair Inglis[2,†], Lucy E. Walker[3,†], Petri Murto[3,†,a], Chiara Delpiano-Cordeiro[1], Colin Morrison[2,5], Naitik A. Panjwani[4], Yao Fu[1,3], Yan Sun[6], Wei Zhou[6], Peter J. Skabara[5], Akshay Rao[1], Alexei Chepelianskii[6]*, Hugo Bronstein[1,3,*], Sam L. Bayliss[2,*], Richard H. Friend[1,*]

[1]The Cavendish Laboratory, University of Cambridge, J. J. Thomson Avenue, Cambridge CB3 0HE, United Kingdom.
[2]James Watt School of Engineering, University of Glasgow, Glasgow, G12 8QQ, United Kingdom
[3]Yusuf Hamied Department of Chemistry, University of Cambridge, Lensfield Road, Cambridge CB2 1EW, United Kingdom.
[4]Berlin Joint EPR Lab, Fachbereich Physik, Freie Universität Berlin, 14195 Berlin, Germany.
[5]School of Chemistry, University of Glasgow, Glasgow, G12 8QQ, UK
[6]LPS, Université Paris-Saclay, CNRS, UMR 8502, Orsay F-91405, France.

[a]current address: Department of Chemistry and Materials Science, Aalto University, Kemistintie 1, 02150 Espoo, Finland

*email: rhf10@cam.ac.uk, alexei.chepelianskii@universite-paris-saclay.fr, sam.bayliss@glasgow.ac.uk, hab60@cam.ac.uk

[†]These authors contributed equally

**Abstract**
We report a family of luminescent alternant diradicals which, at room temperature, support a ground-state spin-triplet, near-unity photoluminescence quantum yields, and optical spin addressability. These diradicals comprise trityl groups meta-linked via pyridyl or phenyl groups, enabling optically bright triplet-to-triplet and singlet-to-singlet transitions. At room temperature, we observe optically detected magnetic resonance in these systems at zero magnetic field and a strong magneto-photoluminescence ($\simeq$10% modulation at 2 mT applied magnetic field). Distinct photoluminescence bands (at $\simeq$630 nm and $\simeq$700 nm) show opposite-sign spin-optical responses, arising from spin-selective intersystem crossing between triplet and singlet manifolds. These bright, all-organic diradicals offer a new set of chemically tunable materials for room temperature spin-optical interfaces, paving the way for application as quantum sensors.

**Main**

Molecular semiconductors have been engineered to give high photoluminescence quantum efficiencies, PLQE, as required for efficient OLED displays. Current technology uses molecules with '*closed shell*' (i.e., spin-0) electronic ground states. Recently however, we and others have developed emissive spin radical materials with '*open shell*' (spin >0) electronic structure, for which it is straightforward to engineer bright optical transitions from ground-state spin-1/2 doublets, thereby opening up a new intersection between systems engineered to enhance molecule-based optoelectronics, and the optical-spin interfaces underlying quantum-sensing platforms. Such bright organic ground-state spins have been realised with trityl radicals based on tris(trichlorophenyl)methyl, TTM, groups coupled to electron donors such as carbazole and triphenylamine where the exciton is an emissive charge transfer state[1–6]. Remarkably these can show close to unity PLQE even for red and near-IR emission. This is achieved by decoupling the electronic structure from vibrational modes that ordinarily cause multi-phonon decay[7].

By coupling an anthracene unit to TTM it is possible to couple a doublet excited state to an energy degenerate triplet exciton, which can generate high-spin quintet or quartet excited states[8]. The sublevel populations in these high-spin excited states can be coherently controlled and read via intersystem crossing, ISC, back to the luminescent doublet state. Diradicals composed of two TTM units coupled using conjugated bridges can also show significant PL, including *meta*-carbazole[9] and acridine-based bridges[10]. We have found close to unity PLQE in a *meta*-fluorene linked TTM diradical[11]. These systems show strong magnetic field modulated luminescence, MPL, at low temperatures and high magnetic fields[9–11]. We have previously found for our *meta*-fluorene bridged diradical, that this is due to magnetic field mediated control of ground state spin population of the triplet and singlet spin manifolds[11] that give PL from the triplet near 630 nm and from the singlet near 700 nm. Low temperature optically-detected magnetic resonance, ODMR and light-induced EPR has been reported on our *meta*-fluorene linked TTM diradical[11] and on phenyl-bridged TTM diradicals[12,13].

Room-temperature optical-spin addressability is useful for quantum sensing and has been observed in a range of '*defect centres*' in crystalline structures, particularly the diamond NV⁻ centre[14–18], as well as in the photoexcited triplet states of pentacene[19] and derivatives[19–21] and green and yellow fluorescent proteins[22–26]. With their distinct electronic structure supporting both triplet-triplet and singlet-singlet fluorescence, organic diradicals could provide a new platform for room-temperature optical spin addressability, however, this functionality has remained challenging[13,27]. Here we report such behaviour with a family of diradical systems that show near-unity photoluminescence quantum yields, and spin-optical addressability at room temperature, demonstrated through both low-field magneto-photoluminescence (MPL, < 2 mT) and optically detected magnetic resonance (ODMR).

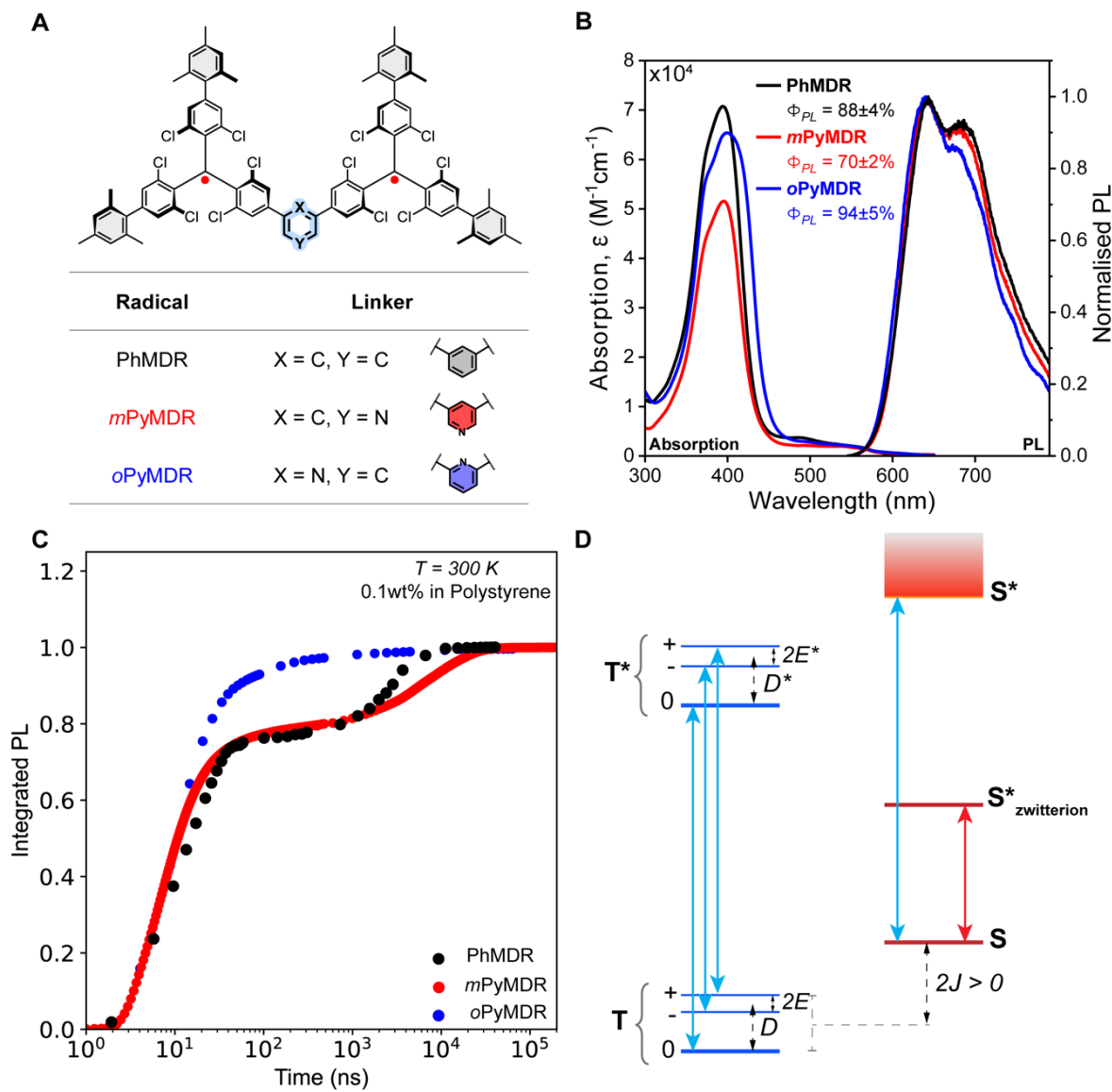

**Figure 1. A family of diradicals with a ground-state spin-triplet and near-unity photoluminescence quantum yield. (a)** Chemical structure of the PhMDR, *m*PyMDR and *o*PyMDR diradicals. The two mesitylated-trityl radical units, $M_2TTM$, are either *meta*-coupled to each other through a phenyl linker or *meta/ortho* connected through a pyridyl linker. The odd-number spacing between the spin-centers favours the spin-triplet (*S*=1) configuration over the spin-singlet (*S*=0) configuration in the ground state. **(b)** Steady-state absorption (solid lines, 100μM toluene solution) and photoluminescence (dashed lines; 0.1wt% of polystyrene thin-film; 405 nm, 10 μJcm$^{-2}$ excitation) for PhMDR (black), *m*PyMDR (red) and *o*PyMDR (blue). The inset shows the PLQEs, which are all close to unity. **(c)** Transient photoluminescence kinetic traces for PhMDR (black), *m*PyMDR (red) and *o*PyMDR (blue). We observe a biexponential PL kinetic for the mPyMDR and PhMDR diradicals with a prompt 9 ns component and a delayed ~10 μs component, the delayed component accounts for >10% of the total PL. **(d)** Proposed Jablonski diagram common to all the diradical systems, comprising a spin-triplet manifolds (T/T*), with zero-field splitting parameters D,E; and a nested spin-singlet manifold (S/S*).

Figure 1.a shows the structure of the diradicals reported here. These are formed by linking dimesitylated TTM, $M_2TTM$, units through a small bridging molecule. The linkage is through the 1,3-positions (*meta*-) of a phenyl linker (PhMDR) or at either the 3,5-positions (*meta*) or 2,6-positions (*ortho*) of pyridyl (*m*PyMDR and *o*PyMDR). Synthesis and characterisation details are provided in Supplementary Section I. Steady-state absorption and photoluminescence, PL, spectra of the different diradicals are shown in Figure 1.b and also Figure S.II.1-3. The three materials show a low oscillator strength

absorption with threshold near 600 nm, involving transitions between the carbon non-bonding level and π and π* levels[4], and excitonic luminescence with a first PL band near 630 nm and a second band near 700 nm.  As we describe below, the 700 nm band is in part a vibrational overtone of the 630 nm band but also has a contribution from a distinct excitonic state[11]. PLQE values are high: in 0.1wt% in polystyrene films the *o*PyMDR diradical has PLQE of 94%, PhMDR  88% and *m*PyMDR 70%. Magnetisation of the powder samples for each diradical was measured using a SQUID magnetometer. The magnetisation plots at 6 T are shown in Figures S.II.3-4.  Ferromagnetic exchange is seen, with deviation of susceptibility from a Curie law below 100K, indicating both a ferromagnetic interaction between the two spins on the diradical, as reported by Kopp *et al*[28] and a ferromagnetic intermolecular exchange.  Very similar behaviour is observed for all three materials.  At room temperature we observe a $\chi T \sim 0.6 \pm 0.1$ emu·K·mol$^{-1}$·Oe$^{-1}$ consistent with a thermalised 1:3 singlet-triplet ratio. Below 10K the magnetisation measured at 6 T saturates at close to 2$\mu_B$ per diradical, the expected value for a 2-spin system with S = 1. Polystyrene thin films show no long range order, and follow the Brillouin magnetisation response expected for S=1, Figure S.II.5-6.  ESR measurements also confirm the presence of ground state triplets, with zero field splitting around 40 MHz for all three materials see Section S.VI.  We note that optically-detected ESR presented below shows a larger triplet level splitting of around 100 MHz, associated with the triplet excited state.  For these *meta*-linked phenyl and pyridyl bridges we note that ferromagnetic exchange ($J > 0$) is expected[12,29–31].

Figure 1.d shows a model Jablonski diagram for the triplet and singlet manifolds expected for these diradicals. We note that both singlet and triplet excitons are optically accessible, and as we develop below, allow spin selective intersystem crossing following photoexcitation. We consider the shorter-wavelength PL at 630 nm is dominated by triplet exciton emission, while the longer-wavelength PL at 700 nm has a larger contribution from singlet exciton emission.  The splitting of the triplet levels, *D* and *E* are set by the separation of the two spins through their dipolar interaction, and are relatively low, no more than ~ 100 MHz, as we show below.

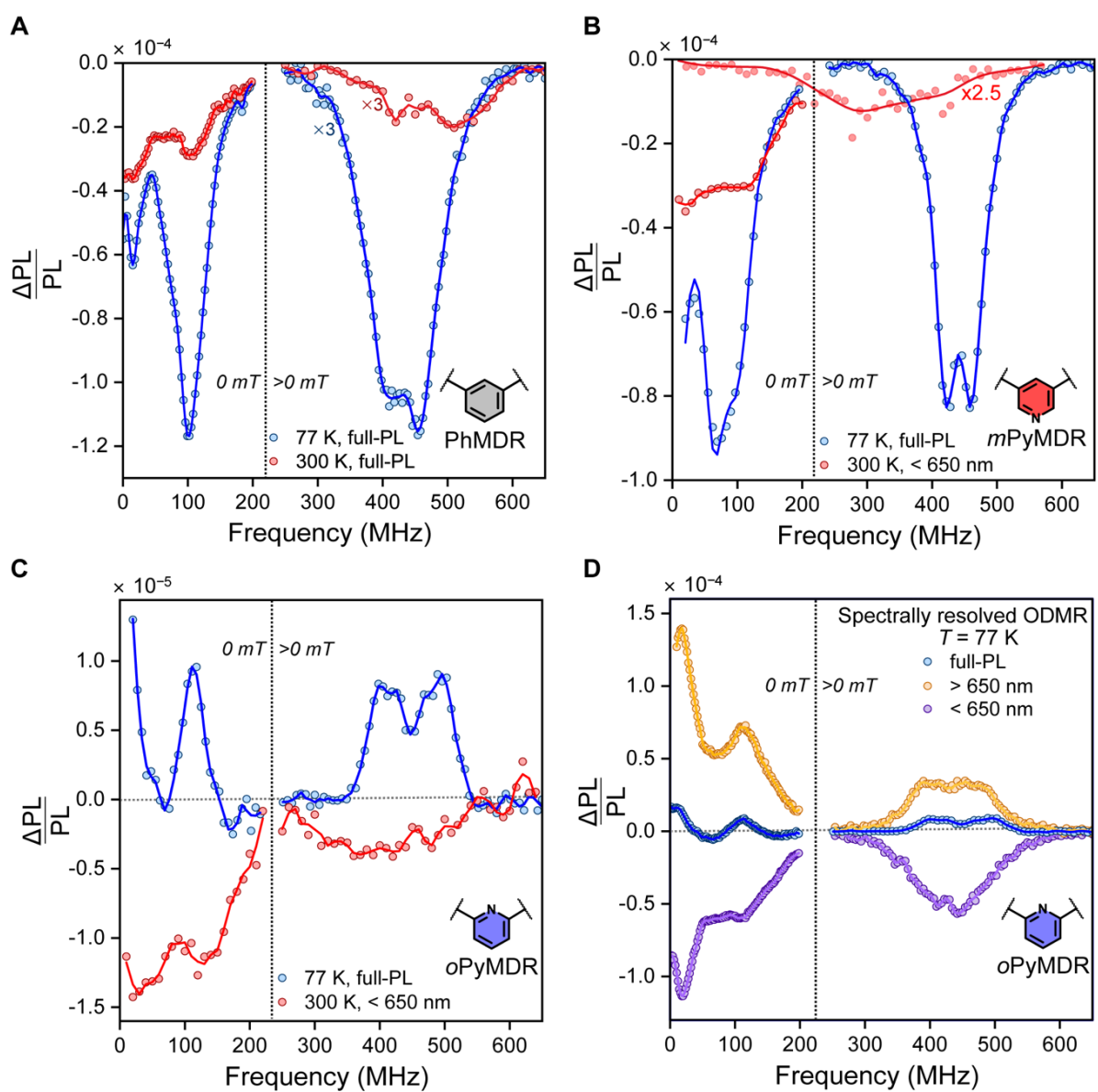

**Figure 2. Room-temperature optically detected magnetic resonance of organic diradicals. (a-c)** Continuous-wave optically detected magnetic resonance, ODMR, spectra at 300 K (red) and at 77 K (blue) under both zero-field and applied magnetic field. **(a)** PhMDR. For applied magnetic field measurements, B=16.75 mT at 300 K and B=15.3 mT at 77 K, **(b)** *m*PyMDR. For applied magnetic field measurements, B=12.2 mT at 300 K and B=15.3 mT at 77K **(c)** *o*PyMDR. For applied magnetic field measurements, B=15.3 mT. Note that 300K measurements for *o*PyMDR and *m*PyMDR were performed using a filter to collect PL photons with wavelength < 650nm, whereas the 77K data were collected for the full PL spectrum. **(d)** Spectrally separated ODMR for *o*PyMDR measured at 77 K at zero-field and 15.3 mT for the full-PL (blue), > 650nm PL (yellow) and < 650nm PL (purple). All experiments used a 0.1wt% of diradical in polystyrene thin-film and 405 nm laser excitation. In all plots the solid-lines, of the same colour as the raw-data points, are median-filtration smoothed raw data lines.

We find that the PL shown in Figure 1.b can be modulated both by microwave resonance (Figure 2) and by small static magnetic fields (Figure 3). Figures 2.a-c show continuous-wave optically detected magnetic resonance, cwODMR, on all three diradicals. We observe ODMR at room temperature and zero applied magnetic field, where the transitions at ≃15 MHz and ≃106 MHz (corresponding to the 2*E* and *D*±*E* transitions within the triplet manifold) yield zero-field splitting parameters of *D*≃100 MHz and *E*≃6 MHz (see Supplementary Section IV for spectral simulations). Decreasing the temperature to 77 K we observe similar transitions as at room temperature, but with an increased ODMR contrast. The larger zero-field splitting parameters observed through ODMR (*D*~100 MHz) compared to ESR (*D*~40 MHz) indicate that the dominant ODMR response is from excited-state triplets (see Supplementary Section IV for details). We note that the ODMR sign is inverted for *o*PyMDR when measuring contrast over the integrated PL, highlighting the chemical tunability afforded with these systems, even under modest chemical modifications. For *o*PyMDR we have presented room temperature ODMR filtering <650nm PL which is of comparable magnitude to the results obtained at 77K. We now discuss the role of spectral filtering in more detail.

The unique electronic structure of these organic diradicals produces an ODMR response which can be enhanced or inverted by collecting only a specific spectral window. Figure 2.d shows the ODMR for *o*PyMDR under three different conditions: *(i)* collecting only PL >650 nm; *(ii)* collecting only PL <650 nm; and *(iii)* measuring all the PL. We find that we can enhance the ODMR contrast by measuring only the shorter-wavelength PL (<650 nm), or invert it, by measuring only the longer-wavelength PL (>650nm) with the contrast for collecting all PL consistent with the sum of these two opposite-sign responses. (We note that the ODMR lineshapes are also slightly different when collecting the different spectral windows - see Supplementary Section IV for details). Importantly, by spectrally filtering, we can significantly enhance the ODMR contrast (8-fold in this case) by separating out competing positive and negative contributions. PhMDR and *m*PyMDR shows a similar spectrally selective behaviour (Figure S.IV.1-2).

To further explore room-temperature optical spin addressability, we measured magneto-photoluminescence, MPL, for the three diradicals. Figure 3.a shows ΔPL/PL, for PhMDR, spectrally filtered between 450 and 650 nm. We observe that spectral filtering increases the MPL contrast by 3-4 times, this is shown in Figure S.III.7-10. The MPL grows with laser fluence, reaching ΔPL/PL = 10% within 2 mT at 50 mW/cm$^2$ with little change at higher fields (verified by measurements up to 9 T, see Figure S.III.3). This laser fluence dependence of the MPL is observed for PhMDR and *m*PyMDR, but is not observed for *o*PyMDR, shown in Figures S.III.7-10. The observed MPL is temperature and excitation-wavelength independent (See Figure S.III.3-4 for more details) for all diradicals, highlighting its robust nature. Figure 3.b-d reports the spectrally resolved PL and MPL at 0 and 5 mT for thin-films of PhMDR, *m*PyMDR and *o*PyMDR. All show larger changes in the 630 nm emission band than for the 700 nm band. With 5 mT applied, PhMDR and *m*PyMDR showed a reduction in the 630 nm peak whereas *o*PyMDR showed an increase.

We suggest a common mechanism governing the ODMR and MPL contrast (see Supplementary Section V for further details) and assign the difference in the peak ODMR contrast (~10$^{-2}$ %) and MPL contrast (~ 10 %) to the weaker spin-mixing efficiency of the microwave rather than static field for the spin ensemble. Given our estimated microwave Rabi frequency of $f_R$<3 MHz and ODMR linewidths of $\Delta f$ ~50 MHz, we expect the contrast to be reduced by $\sim \frac{f_R}{\Delta f} \sim 6 \times 10^{-2}$ compared to the MPL, where the static field can fully mix the spin states, in line with our observations and the lack of ODMR saturation with applied microwave power (see SI).

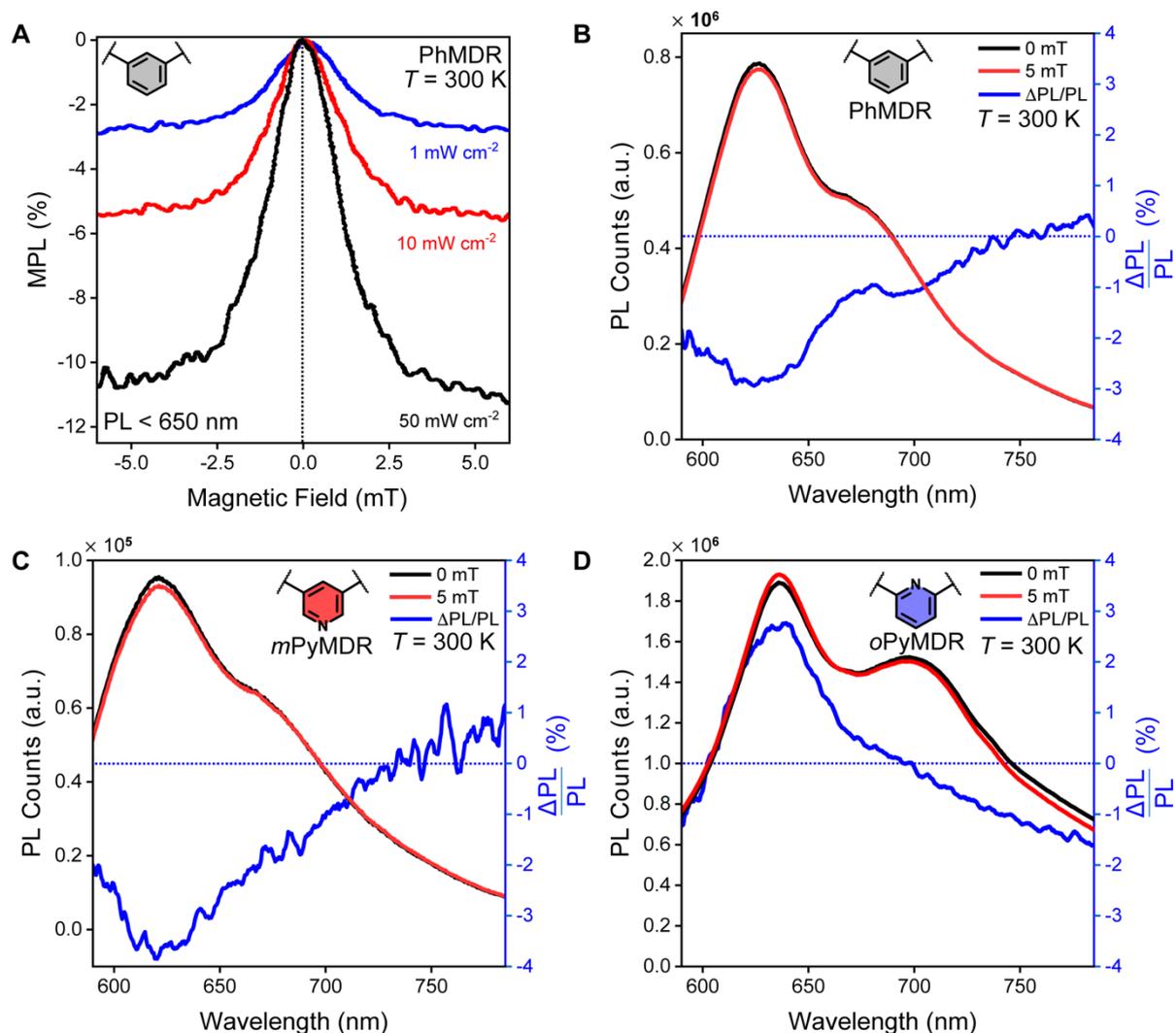

**Figure 3. Magnetic field modulated luminescence. (a)** Magnetic field dependence of the PL intensity in the <650 nm region and in the ±5mT range measured for PhMDR at different laser excitation fluences of 1 mWcm$^{-2}$ (blue), 10 mWcm$^{-2}$ (red) and 50 mWcm$^{-2}$ (black). PL spectra recorded at 0mT (black) and 5mT (red) and the difference spectrum (blue) for **(b)** PhMDR, **(c)** *m*PyMDR and **(d)** *o*PyMDR. Samples were thin films of 0.1wt% of diradical in polystyrene. 405 nm laser excitation with an excitation density of 1 mWcm$^{-2}$ was used for data shown in panel (a). Laser excitation at 400 nm at an excitation density of 1 mWcm$^{-2}$ and a repetition rate of 10 kHz was used for data shown in (b)-(d).

Transient photoluminescence, TrPL, studies, shown in Figure 1.c, show that upon photoexcitation PhMDR and *m*PyMDR has a prompt component peaked at 640 nm with a lifetime of ~10 ns (see Figure S.VII.3 for spectrally resolved data). Within the first 50 ns the PL evolves into a 700 nm-peaked spectrum, with more than 10% from beyond 100 ns, and this persists out to 100 µs. For *o*PyMDR, we obtain PL peaking at 640 nm with no spectral evolution over time, and a dominant single decay having a lifetime of 9 ns. Spectrally-resolved transient absorption, TA, are shown in Figure 4.a-c for the three diradicals. All show a broad excited state absorption band from 450 to beyond 600 nm, with lifetime of order 10 ns, that clearly tracks the initial decay of the PL. At longer times, PhMDR and *m*PyMDR show red-shifted absorption bands, with peaks at 600nm and 750nm that persist to beyond 100 µs see figure S.VII.4. We note that these spectral features are consistent with anionic and cationic absorptions centred on trityl units[32]. In contrast, *o*PyMDR shows little evidence for this long-lived band. We consider these delayed features are due to long-lived '*zwitterionic*' singlet excitons where one electron has been transferred from one radical site to the other[11].

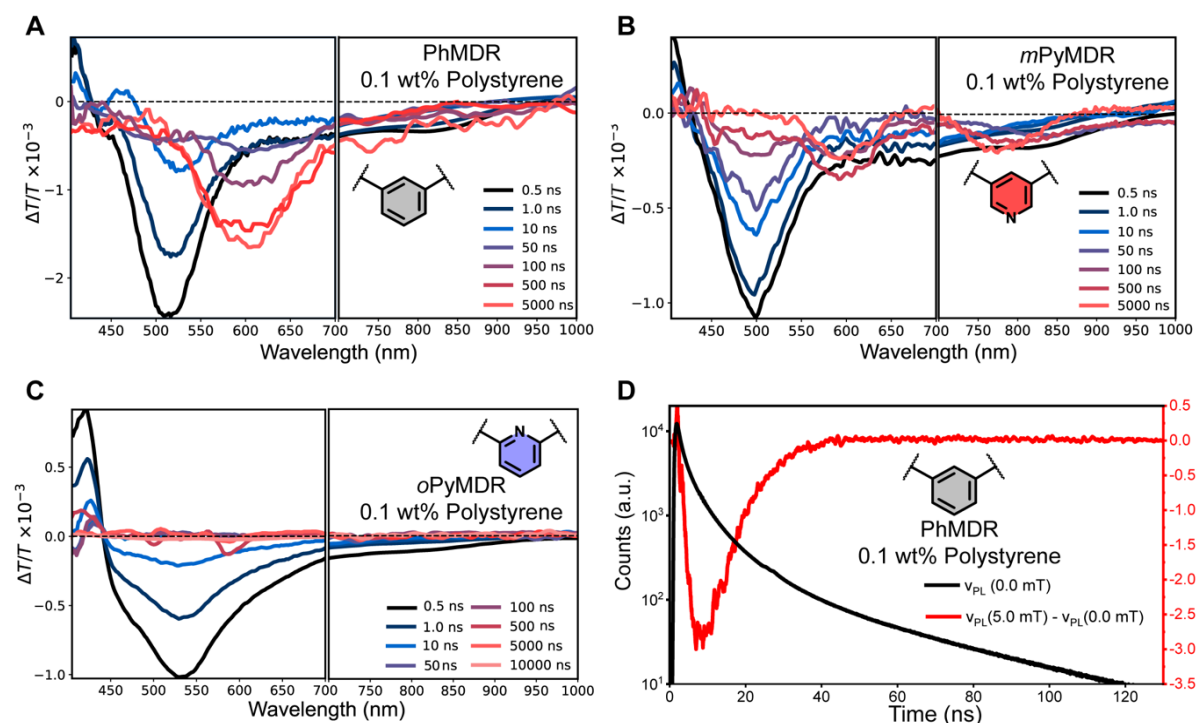

**Figure 4. Photophysics of diradicals.** Transient absorption, TA, spectral slices of **(a)** PhMDR, **(b)** *m*PyMDR and **(c)** *o*PyMDR with positive features corresponding to ground-state bleach, GSB, and negative features corresponding to excited state absorptions, ESA. **(d)** Magnetic field dependent change in PL counts resolved in time, we show the kinetic trace at 0 mT (black) and the time-resolved change in counts at 5 mT (red). Samples were thin films of 0.1wt% of diradical in polystyrene. Laser excitation at 400 nm and a fluence of ~5µJcm$^{-2}$ was used.

Evidence for spin-selective intersystem crossing in the excited state can be obtained from transient PL measurements. Figure 4.d shows the change in PL with 5 mT applied magnetic field. This shows that the magnetic field induced PL quenching builds up during the first 9 ns, this is within the lifetime of the prompt luminescence for PhMDR, the built-up PL-quenched states decay over the following 30 ns. This growth is dependent on the applied field, as shown in Figure S.III.6. We consider that this field-dependent kinetics change arises from a spin-sublevel selective intersystem crossing as elaborated further in Supplementary Section V.

The singlet and triplet manifolds illustrated in Figure 1.d provide the electronic structure for spin-selective intersystem crossing that can account for the MPL and ODMR we report here. Figure S.III.1 shows that on freezing out of the singlet exciton population (25% at 300K) at low temperatures, the 700 nm band falls by 20%. We consider that triplets decay radiatively, principally in the 630 nm band. There are more options for singlet excitons. They can also decay radiatively but in competition with both ISC to the triplet and also relaxation to the lower energy zwitterionic charge-transfer exciton, identified in Figure 1.d, which we consider is responsible for the long-lived excited states shown in Figure 4. This scheme may allow contributions from both excited state and ground state polarisation.

The three materials we report here show unexpectedly high luminescence efficiencies (> 75% at 300K), and contrast these and other mesityl-substituted diradicals[11] with other reported diradicals where efficiencies are generally much lower at room temperature. Our findings also demonstrate low magnetic field and microwave control at room temperature and provide a new family of highly emissive materials that can be used for room-temperature spin-optical interfaces. The canonical example of a room-temperature spin-optical interface is the diamond nitrogen vacancy, NV-, centre which has been developed for a large variety of quantum sensing applications[17,33–40], with progress being made on other defect systems[41–48], but with a molecular platform opening new opportunities for precise spatial control, nanoscale integration, and ability to work at high spin densities due to well-defined host-guest chemistry.

We note that there is interest in the use of this class of materials for biosensing; for example as developed using 'nanodiamond' NV- centres to enhance detection limits in 'lateral flow' tests[17]. We consider here the potential performance of our molecular sensors against nanodiamonds in this application, as a point of comparison. Our systems show several attractive characteristics: *(i)* these diradicals comprise relatively well separated radical sites, causing the dipolar coupling, *D* to be small, at around 100 MHz, in comparison to the GHz values for literature systems such as the diamond NV centre, giving high MPL response at very low B fields. MPL is thus particularly straightforward to measure, and may have advantages over microwave modulation methods. Magnetic fields for the DC measurements shown in figure 3 were provided by a Helmholtz pair, each copper coil with 1500 turns on a former of diameter 2 cm, to give 0.1mT/mA. We also drove these coils to give AC fields at a few hundred Hz and detected the same MPL response at twice the modulation frequency. *(ii)* it is straightforward to form guest-host composites using non-polar polymer hosts. This is exemplified here using polystyrene to fabricate thin films. We find these guest-host systems show good stability under ambient experimental conditions – results reported in Figures 2-3 were obtained with no further encapsulation after the films were cast on the substrate, see Figure S.II.4 for an image of a thin film sample under photoexcitation. *(iii)* we are able to achieve relatively high concentrations of active fluorophores while retaining full PLQE and MPL contrast, at weight concentrations of 1%, see Figure S.III.7. This gives 2500 molecular emitters in a 100 nm sphere, and an effective improvement in emitter density over nanodiamonds by a factor of 10-100.

Looking forward, we consider that our materials establish a new class of molecular spin-optical materials. They offer the promise for further synthetic control, such as direct attachment to biological targets. There is scope to tune emission across the '*biological window*' to 800 nm, as demonstrated for monoradicals[49]; and scope to develop bright multi-spin architectures. Their properties enable convenient luminescence modulation using very low magnetic fields, at levels comparable to the spin radical pairs associated with magnetic field navigation in songbirds[50,51].

**Acknowledgements**

We acknowledge Sebastian Gorgon, Biwen Li and Lujo Matasovic for useful discussions. **R.H.F**, **L.E.W, P.M.** and **R.C.** received funding from the European Research Council under the European Union's Horizon 2020 research and innovation programme (Grant Agreement No. SCORS – 101020167). **H.A.B** acknowledges support from the Engineering Physical Sciences Research Council (EPSRC, grant number EP/S003126/1). **A.C** and **Y.S** acknowledge support from ANR-20-CE92-0041 (MARS), IDF-DIM SIRTEQ, and the European Research Council (ERC) under the European Union's Horizon 2020 research and innovation programme (grant Ballistop agreement no. 833350). **Y.F** thanks financial support by the Engineering Physical Sciences Research Council (EPSRC, grant number EP/W017091/1) Programme Grant. **A.R** received funding from the European Research Council under the European Union's Horizon 2020 research and innovation programme (Grant Agreement No. 758826). **S.B, A.I** and **C.M.** acknowledge support by UK Research and Innovation [grant number MR/W006928/1]

**Author Contributions Statement**
**R.H.F, R.C.** and **H.A.B** conceived the project. **R.C** developed the project, planned and conducted the steady-state spectroscopy, transient spectroscopy and magneto-optical experiments. **L.E.W** and **P.M** designed, synthesized the molecules, characterised the synthesized molecules, assessed their purity and carried out the cyclic voltammetry studies. **A.I**, **C.M.** and **S.B.** performed and analysed ODMR measurements. **P.S** supervised **N.A.P.** and **Y.F.** performed the CW, and **N.A.P.** Pulsed ESR. **Y.S**, **W.Z.** and **A.C.** performed and analysed the cryo-magneto-optical experiments. **C.D.C.** performed photophysical measurements. **R.H.F, R.C, H.A.B, A.R.** and **S.B.** wrote the manuscript with contributions from all other authors.

## Methods
**Characterization and techniques.** NMR spectra were recorded on a 400 MHz Bruker Avance III HD spectrometer ($^1$H, 400 MHz; $^{13}$C, 100 MHz) or 500 MHz Brucker DCH Cryoprobe Spectrometer ($^1$H, 500 MHz; $^{13}$C, 126 MHz). Chemical shifts are reported in $δ$ (ppm) relative to the solvent peak, chloroform-*d* ($CDCl_3$: $^1$H, 7.26 ppm; $^{13}$C, 77.16 ppm) and dichloromethane-*d*$_2$ ($CD_2Cl_2$: $^1$H, 5.32 ppm; $^{13}$C, 53.84 ppm). Mass spectra were obtained on a Waters Xevo G2-S benchtop QTOF mass spectrometer equipt with a electrospray ionization (ES) or atmospheric solids analysis probe (ASAP). C, H, N combustion elemental analyses were obtained on an Exeter Analytical Inc. CE-440 elemental analyzer and the results are reported as an average of two samples. Flash chromatography was carried out using Biotage Isolera Four System and Biotage SNAP/Sfär Silica flash cartridges.
**Steady state optical absorption and photoluminescence measurements.**
Ultraviolet/visible/NIR spectra were measured with a commercially available Shimadzu UV-2550 spectrophotometer and a Shimadzu UV-1800 spectrophotometer.

Photoluminescence was measured in a home-built setup by providing a continuous wave excitation at 532 nm using a diode laser. Photoluminescence is collected in a reflection mode setup after passing photons through a 550 long-pass filter (Thor Labs). The transmitted photons then are collected in a collimating 2-lens apparatus and directed into an optical fiber which supplies the photons into a calibrated grating-spectrometer (Andor SR-303i) and finally into a Si-camera where it is recorded. Output spectra are corrected taking into account the filter transmission and camera sensitivity. The excitation spectra were measured with a commercially available Edinburgh Instruments FS5 Spectrofluorometer system using a xenon lamp light source. When solution samples are studied 100 mm pathlength Hellma quartz cuvettes were used. In all cases where thin-film samples were studied, we used un-encapsulated 0.1wt% of diradical in polystyrene thin-films with thickness ~1 μm.

**Cyclic Voltammetry (CV).** CV measurements were carried out on a PalmSens EmStat4S potentiostat in a three-electrode setup using a glassy carbon electrode (3.0 mm diameter) as the working electrode (WE), platinum wire as the counter electrode (CE) and freshly activated silver wire as the $Ag/Ag^+$ reference electrode (RE). The silver wire was activated by immersing in concentrated HCl solution to remove any silver oxides and other impurities, then rinsed with water and acetone and dried prior to each measurement. The RE was calibrated against ferrocene/ferrocenium ($Fc/Fc^+$) redox couple at the end of each measurement. The $Fc/Fc^+$ half-wave potential, $E_{1/2}$, was determined at 0.20 V vs. $Ag/Ag^+$ in THF electrolyte and at 0.50 V vs. $Ag/Ag^+$ in DCM electrolyte. The supporting electrolyte was 0.1 M solution of $Bu_4NPF_6$ in THF (anhyd.) and in DCM (anhyd.). The scan rate was 0.1 V s$^{-1}$. The electrolyte was bubbled with Ar gas before each measurement to remove any dissolved oxygen. Sample concentration was in the order of $10^{-5}$ M.

**Photoluminescence Quantum Efficiency (PLQE).** Steady-state PLQE measurements were performed using an integrating sphere. A continuous-wave 405 nm excitation is provided by a 405 nm diode laser (Thorlabs LP405C1) with excitation powers of 10–30 mW cm$^{-2}$. A focussed beam of diameter 1000 μm was used to excite the samples. The emission was directed using an optical fiber in a calibrated grating spectrometer (Andor SR-303i) onto a Si-camera.
Repeat measurements of PLQE were performed independently using an Integrating sphere on an FLS-1000 Photoluminescence Spectrometer with an integrating sphere attachment and on another integrating sphere where a continuous-wave 405 nm excitation is provided by a laser diode (Thorlabs DL5146-101S). The emission was collected in an integrating sphere and directed into an Andor iDus Si camera. Calculations were based on 3 configurations – empty sphere, sample perpendicular to laser beam and sample parallel to laser beam as described by DeMello *et al* [52].

**Time Correlated Single Photon Counting (TCSPC).** The studied solution samples are irradiated using an electrically pulsed 405 nm laser using a function generator at a frequency of 20 MHz providing a time resolution of upto 100 ns. Photons emitted from the sample were passed through a 405 nm long-pass filter (Thor Labs Ltd.) to remove the laser scatter. The subsequently transmitted photons are then collected by a Si-based single-photon avalanche photodiode. The instrument response function was found to be ~0.1 ns in this setup.

**Continuous-Wave Optically Detected Magnetic Resonance (cw-ODMR).** For cw-ODMR, samples were excited using a 405 nm laser diode (ThorLabs LP405-SF10), which was filtered using a 450 nm short-pass filter (ThorLabs FESH0450). A dichroic mirror (ThorLabs DMLP425R) was used to direct the excitation laser to a 1" achromatic doublet lens (ThorLabs AC254-030-AB) which was used to focus the light into a 400-micron diameter core optical fibre (FT400EMT) that was placed in contact with the sample. The photoluminescence was collected by the same fibre, then collimated via the lens, separated from the laser excitation via the dichroic, and passed through a 550 nm long-pass filter (ThorLabs FELH0550) before being focussed onto a free-space detector (FEMTO, OE-200-SI) using a second achromatic doublet (ThorLabs AC254-030-AB). For spectrally selective measurements (Figure 2, d), either a 650 nm longpass filter (Thorlabs FELH0650) or a 650 nm shortpass filter (Thorlabs FESH0650) were added to the collection path. For 77K measurements, the sample was submerged in liquid nitrogen in a dewar.

The microwave fields were applied using a co-planar waveguide connected to a signal generator (Agilent E8257D). The microwaves were square-wave modulated using a microwave switch (Mini-circuits ZASWA-2-50DRA+) with a modulation rate of 487 Hz, and amplified using a high-power microwave amplifier (Mini-circuits LZY-22+ , Mini-circuits ZHL-25W-272+, or ZHL-5W-202-S+). The PL signal from the free-space detector was passed into a lock-in amplifier (Stanford Research Systems, SR830) which was referenced to the microwave modulation frequency. The lock-in signal yielded the ODMR signal ΔPL=PL(Microwaves on) - PL(Microwaves off) which we normalize to the DC PL which we measure using an auxiliary input to the lock-in amplifier.

To apply a static magnetic field $B_0$ we used a neodymium permanent magnet (Magnosphere, 1373) secured in a 3D printed rig. The magnet was placed underneath the sample with the field orientation parallel to the optical axis. The magnetic field strength was calibrated by performing ODMR on a NV centre diamond spin ensemble. Room temperature measurements of *m*PyMDR and *o*PyMDR were performed using an electromagnet (GMW Magnet Systems Model 3470) driven by a bipolar current supply (KEPCO, BOP 50-8) to apply a static magnetic field perpendicular to the microwave field. An 3D magnetic sensor field (Infineon, TLE493D-W2B6 MS2GO) was used to calibrate the magnetic field. The optical set up was as described above for the 77K measurements without the optical fibre in contact with the sample. In this case, the excitation laser was focused onto, and PL collected from the sample.

**Transient Photoluminescence (TRPL) Spectroscopy.** Transient photoluminescence spectra at nanosecond-microsecond timescales were recorded using an electronically gated intensified CCD (ICCD) camera (Andor iStar DH740 CCI-010) connected to a calibrated grating spectrometer (Andor SR303i). A narrowband non-colinear optical parametric amplifier pumped with a frequncy doubled output of a 1 kHz 800 nm laser pulse from a Ti:sapphire amplifier was used to generate a tunable 250-fs excitation pulse. Suitable long-pass filters (Edmund Optics) were used to prevent scattered laser signals from entering the spectrometer. Temporal evolution of the PL emission was obtained by stepping the ICCD delay with respect to the excitation pulse, with a minimum gate width of 5 ns. The raw data was corrected to account for filter transmission and camera sensitivity.

In the same setup we measured magnetic field effects on the photoluminescence and electroluminescence. where the sub-1T magnetic field was generated using an electromagnet from GMW Model 3470 with 1 cm distance between cylindrical poles and the field strength calibrated with a Gaussmeter.

**Transient Absorption (TA) Spectroscopy.** Transient absorption experiments were conducted on a setup pumped by a regenerative amplifier (Light Conversion, Pharos) emitting sub-200 fs pulses centered at 1035 nm at a rate of 10 kHz. The output of the amplifier was optically delayed up to 8 ns by a multi-pass computer-controlled delay stage. This was used to generate a 500–1000 nm white light (WL) in a sapphire crystal. For generating a 350–500 nm WL, the seed was doubled in a BBO crystal before the sapphire. Wavelength-tunable pump pulses were generated in an Orpheus Neo (Light Conversion) unit. The pump was chopped at 100 Hz to provide on average a 5 kHz repetition rate for pump-on and pump-off measurements. The pump spectrum was filtered using appropriate band-pass filters to remove residual wavelengths, and its polarization set to magic angle relative to the probe pulses using a Berek rotator. The pump and probe beams were spatially overlapped at the focal point using a beam profiler. The pump and probe diameter at the sample position were on the order of 1000 and 100 µm respectively. After passing through the sample, the probe beam was dispersed with a grating spectrometer (Kymera, Andor Technology) and measured with a Si detector array.

**Global Analysis.** Global analysis is performed using the sum of weighted components analysis where the spectrum is, at a first approximation, expressed as a *i*-term linear sum of spectral components $\varphi_i(\lambda)$:

$$f(\lambda,t) = \sum_i \varphi_i(\lambda) e^{-t/\tau_i}$$

Each $i^{th}$ spectral component is weighted by a time-varying weight pre-factor computed from its associated lifetime $\tau_i$ given by $e^{-t/\tau_i}$. This equation can be cast into a n x m matrix using the method described by Dorhliac et al[53]. This is a matrix consisting of n-reduced weight singular vectors (represented by $(US)_n$) that can completely reconstruct the experimental data matrix. The matrix formalism equation is then written as

$$(US)_n = E(\vec{\tau})_{n\times m} X_{m\times 1}$$

Where the time-varying exponential weight matrix is given by $E(\vec{\tau})$ and $X$ is the wavelength dependent weight matrix, the dimensions of each term is given in the subscript. A simulated annealing algorithm[54] used for global minimization routines is then used to determine which lifetimes $\tau_i$ can best satisfy the matrix equation. The determined lifetimes from the optimization are then inserted into the matrix equation and then it is solved for X by solving the least-square problem[55]:

$$min_\tau |(US)_n - E(\vec{\tau})_{n\times m} X_{m\times 1}|^2$$

**Magneto-Photoluminescence measurements.** Magneto photoluminescence (MPL) measurements were carried out on Bluefors dry dilution refrigerator with a superconductive vector magnet from American Magnetics Model. The samples were installed on a gold-coated copper plate and cooled down to 250 mK. The excitation source was a 405 nm (3.06 eV) laser stabilized through a PID loop, which was also used to provide power amplitude modulation at 1.2 kHz. A fiber was used to guide the excitation to the sample (excitation power of around 20 nW, 50 µW/cm$^{-2}$) and to collect the fluorescence signal. The magnetic field was swept from 0 to 9 T with a ramping speed of 0.1 Tesla per minute. MPL variation at a particular wavelength, was collected using an avalanche photodiode (Thorlabs) placed behind a monochromator (HORIBA iHR320) with lock-in detection at the power amplitude modulation frequency.

For room temperature studies the sample holder was kept at ambient temperature and fixed inside a superconducting solenoid providing a variable static magnetic field ($B_0$). As described earlier, one optical fiber was used to guide the laser into the sample stage and another fiber was used to collect PL. Any MPL or M-TCSPC variation at a particular wavelength, was collected using an avalanche photodiode (Thorlabs) or a fast PicoQuant photodiode placed behind a monochromator (HORIBA iHR320) with lock-in detection at the laser power amplitude modulation frequency.

For the DC measurements mentioned in the paper we used a Helmholtz coil pair. Each coil of the pair was constructed by winding a copper coil 1500 times on a former of diameter 2 cm, we were able to produce 0.1mT of magnetic field for every 1 mA applied across the coil. For the AC measurements one of these coils were used, a 3D printed holder was fabricated to hold the sample being studied at the center of the coil.

**Continuous wave ESR (cwESR).** cwESR spectra were recorded at X-band frequencies (~9.4 GHz) using a laboratory-built ESR spectrometer. The magnetic field was regulated using a Bruker BH15 field controller and monitored with a Bruker ER 035M NMR Gaussmeter while a Bruker ER 041 MR microwave bridge (with a ER 048 R microwave controller) was used for microwave generation and detection (diode-detection). The static magnetic field was modulated at 100 kHz and lock-in detection was carried out using a Stanford Research SR810 lock-in amplifier in combination with a Wangine WPA-120 audio amplifier. The lock-in detection leads to the derivative spectra. A Bruker ER 4122 SHQE microwave resonator was used. g-factor calibration was additionally done using a N@C60 powder sample at room temperature, with a known g-factor (g =2.00204). The inner wall film sample of PhMDR, *m*PyMDR and *o*PyMDR was measured at room temperature with a modulation amplitude of 0.05 mT and a microwave power of 3.2 µW for the main transition and 12.6 mW for the half field transition. Simulations of the cwESR spectrum were done using the EasySpin toolbox[56], version 6.0.0-dev.53 and the pepper function. The half field transition was scaled to account for the difference in microwave power used.

**Pulsed ESR.** Pulse ESR was performed at X-band (9.7 GHz) using a Bruker ELEXSYS E580 spectrometer, with a Bruker MD5 resonator. All measurements were at room temperature in the dark.

The echo-detected field sweep was performed using the Hahn echo sequence $\frac{\pi}{2} - \tau - \pi - \tau -$ echo, with pulse lengths of $t_{\pi/2} = 128$ ns and $t_\pi$ = 256 ns, an inter-pulse delay τ of 400 ns, a two-step phase cycle and an echo integration window of 500 ns. The longer pulses were used to supress ESEEM effects.

The Rabi nutation trace was recorded with a three-pulse sequence of θ-4µs-π/2-τ-π-τ-echo where θ is a variable length pulse which is incremented. The nutation trace was recorded with two different π/2 and π pulse lengths 1) where $\tau_{\pi/2}$ = 128 ns and $\tau_\pi$ = 256 ns, and 2) $\tau_{\pi/2}$ = 16ns and $\tau_\pi$ = 32 ns.

The $T_M$ (phase memory) time was measured with an echo decay measurement with a pulse sequence π/2-τ-π-τ-echo where τ was incremented from 400 ns to 2.8 µs in steps of 4 ns. $\tau_{\pi/2}$ = 128ns and $\tau_\pi$ = 256 ns.

The $T_1$ (spin-lattice) relaxation time was measured with an inversion recovery pulse sequence, with π-T-π/2-τ-π-τ-echo where T is incremented from 280ns to 75.28ms. $t_{\pi/2}$ = 16ns and $t_\pi$ = 32ns with a integration window of 98ns.

Field calibration was carried out using N@C60.

**Magnetization measurements.** Powder samples were encapsulated and mounted in a gelatin capsule and inserted into a straw sample holder, the ends were held in place using non-magnetic tape, with the sample position being ~66 mm in the sample area within the magnetometer. The DC-magnetisation was measured by warming on a Quantum Design MPMS®3 in the temperature range 2.0 – 300 K under an applied magnetic field of 60000 Oe (6T), after cooling the sample from 300 K to the base-temperature (2-4 K) in zero field (ZFC). The isothermal magnetisation was measured on the same system in the field range µ₀H = 0–6 T at the lowest temperature obtained (4K).